\documentclass[12pt]{article}

\usepackage{a4wide}
\usepackage[ngerman,english]{babel}
\usepackage{amsmath, amssymb, latexsym}
\usepackage{graphicx}
\usepackage{bm}
\usepackage{hyperref}
\usepackage{color}
\usepackage[dvipsnames]{xcolor}
\usepackage{soul}
\usepackage{cancel}

\begin{document}

\begin{titlepage} 

\begin{flushright}
{\small{}LMU-ASC 60/18} 
\par\end{flushright}

\begin{center}
\vspace{1mm}

\par\end{center}

\begin{center}
\textbf{ \Large{}Bulk-boundary correspondence between charged, anyonic strings and vortices}
\par\end{center}
\vspace{1mm}
\begin{center}
Alexander Gu{\ss}mann$^{1}$, Debajyoti Sarkar$^{2}$ and Nico Wintergerst$^{3}$
\\
 
\par\end{center}
\vspace{1mm}
\begin{center}
$^{1}$\textsl{\small{}Arnold-Sommerfeld-Center for Theoretical Physics},\\
\textsl{\small{} Ludwig-Maximilians-Universit\"at, Theresienstr. 37,
80333 M\"{u}nchen, Germany}\\
$^{2}$\textsl{\small{}Albert Einstein Center for Fundamental Physics, Institute for Theoretical Physics,}\\
\textsl{\small{}  University of Bern,
Sidlerstrasse 5, CH-3012 Bern, Switzerland}\\
$^{3}$\textsl{\small{}The Niels Bohr Institute, University of Copenhagen},\\
\textsl{\small{} Blegdamsvej 17, 2100 Copenhagen \O,
Denmark}\\
\par\end{center}{\small \par}

\vskip 0.3 cm
\begin{abstract}

We discuss a unified framework of dealing with electrically charged, anyonic vortices in 2+1 dimensional spacetimes and extended, anyonic string-like vortices in one higher dimension.
We elaborate on two ways of charging these topological objects and point out that in both cases the vortices and strings obey fractional statistics as a consequence of being electrically charged.
The statistics of the charged vortices and strings can be obtained from the phase shift of their respective wave-functions under the classic Aharonov-Bohm type experiments.
We show that for a manifold with boundary, where one can realize 2+1 dimensional vortices as endpoints of trivially extended 3+1 dimensional strings, there is a smooth limit where the phase shift of a bulk string-vortex goes over to the phase shift of the boundary vortex. This also enables one to read off the bulk statistics (arising essentially from either a QCD theta-type term or an external current along the string) just from the corresponding boundary statistics in a generic setting. Finally, we discuss various applications of these findings, and in particular their prospects for the AdS/CFT duality.
\end{abstract}
\end{titlepage}

\section{Introduction}

The physics of fractional statistics and vortices is well studied in quantum field theory. 
If the positions of two identical point particles 
in $(3+1)$ (or in higher) spacetime dimensions are interchanged, the corresponding wave function acquires a multiplicative factor of either $(+1)$ (Bosons) or $(-1)$ (Fermions). The spin of the point particles in $(3+1)$ takes either integer values (for Bosons) or half-integer values (for Fermions) \cite{Pauli:1940zz}. 
However, in \cite{Wilczek:1982wy} it was shown that in $(2+1)$ spacetime dimensions point particles can also carry an arbitrary spin and can obey fractional statistics. These point particles with fractional spin and statistics are known as ``anyons". If an anyon is transported around another identical anyon, the combined wave function acquires a factor of $\exp\left[{\pm\, i\frac{\alpha}{\pi} \phi}\right]$ where $\phi$ is the angle of rotation and all real values of the parameter $\alpha$ can be realized. Also, $\pm$ sets the convention of which way is it rotated. In $(2+1)$ dimensional quantum field theories such anyons can be realized in certain cases if a Chern-Simons term is present in the field theory Lagrangian \cite{Wilczek:1983cy, Arovas:1985yb}. One particular well known example is the case of electrically charged Chern-Simons vortices: according to the theorem of Julia and Zee \cite{Julia:1975ff}, finite energy $(2+1)$ dimensional vortices of Nielsen-Olesen type \cite{Nielsen:1973cs} cannot be electrically charged. If however a Chern-Simons term is added to the Nielsen-Olesen Lagrangian, electrically charged vortices do exist as $(2+1)$ dimensional topologically non-trivial static lowest energy configurations in this theory \cite{Paul:1986ix, deVega:1986eu} and these electrically charged vortices can obey fractional statistics \cite{Frohlich:1988qh}.
\newline

In higher spacetime dimensions, point-like objects cannot obey fractional statistics because their corresponding braid group is trivial.
 However, in $n$ space dimensions, the (higher dimensional analogue of the) braid group for ($n-2$)-dimensional objects is non-trivial \cite{Manin}, allowing for the interesting possibility of field theories with such objects obeying fractional statistics. For example, there is the possibility that string-like objects of certain quantum field theories can obey fractional statistics in $(3+1)$ spacetime dimensions, as discussed e.g. in \cite{Aneziris:1990gm,Aneziris:1992aj} in a concrete setup. For works on anyonic strings and membranes in string theory and in particular in AdS, see e.g. \cite{Hartnoll:2006zb}. For applications of similar theories towards AdS/condensed matter see e.g. \cite{Roychowdhury:2014gja}. 
\newline

In this work, we investigate this topic in the context of a bulk manifold $\mathcal{M}$ with a boundary $\partial\mathcal{M}$.
The endpoints of string-vortices in $\mathcal{M}$ can be understood as point-like vortices (or antivortices) located on $\partial\mathcal{M}$.\footnote{It is not guaranteed that string-vortex solutions exist for any $\mathcal{M}$. We are obviously confining our attentions to spacetimes where such a solution is available. A typical example will be when $\mathcal{M}$ is AdS, which has a conformal boundary and also supports a string-vortex solution \cite{Dehghani:2001ft}.}
As we shall discuss, the statistics of such strings finds a direct interpretation in terms of the statistics of the boundary vortices. 
For this purpose we consider Abelian `cosmic' strings in a curved $(3+1)$ dimensional spacetime $\mathcal{M}$, which end on the $(2+1)$ dimensional  boundary $\partial\mathcal{M}$ and investigate the question if and under what conditions these cosmic strings obey fractional statistics. We argue that under quite general conditions, one can answer this question by considering only the endpoint vortices of the cosmic string. 
\newline

The conceptual idea behind our work is rather straightforward. The statistics of vortices in $(2+1)$ dimensions is captured by a low energy effective theory whose main ingredient is a Chern-Simons term. The presence of the Chern-Simons term attaches magnetic flux to electric charges. It is this combination of flux and charge that gives rise to non-trivial Aharonov-Bohm (AB) phases when two such topological object are rotated around each other. A simple continuation to one additional space dimensions involves the presence of either $\theta$-like terms or mixed Chern-Simons-like terms. This in turn can induce fractional statistics of 'cosmic' strings through corresponding AB phases. This mechanism, in fact, may appear somewhat unconventional from the bulk perspective, where the topological actions of Chern-Simons type usually dictate that we perform such an AB measurement in a $(4+1)$ dimensional bulk \cite{Hartnoll:2006zb}.
\newline

Note that, \emph{both} in the $(2+1)$ dimensional boundary and $(3+1)$ dimensional bulk, there are subtleties related to self-linking and intersections of their corresponding worldlines or worldsheets \cite{Aneziris:1992aj,Bergeron:1994ym,Polikarpov:1993cc}. To avoid such complications, we will take the strings to be far separated such that they are always parallel and are trivially extended from the boundary charged particle vortex.
 As we will argue, in this setting, cosmic strings obey fractional statistics if (i) their endpoint vortices are electrically charged and (ii) charges are arranged such that the Aharonov-Bohm phase of a vortex pair is not canceled by that of the antivortex pair corresponding to the opposing endpoints of the strings. This can be achieved in several ways, as we shall discuss in the bulk of the paper.
\newline

In order to illustrate our points in the $(3+1)$ dimensional case, we investigate two particular setups. First, we study the case of Abelian cosmic strings in $\mathcal{M}$ which are obtained as finite energy configurations in the theory with Nielsen-Olesen Lagrangian with a term $\Delta \mathcal{L} = \theta \epsilon^{\mu \nu \alpha \beta} F_{\mu \nu} F_{\alpha \beta}$ added. This is of course the familiar `$\theta$-term' like quantity of QCD, although the field strength $F_{\mu\nu}$ now corresponds to a $U(1)$ gauge field $A_{\mu}$. We consider a setup with these Abelian cosmic strings, with both endpoints of the string ending on the $(2+1)$ dimensional boundary of $\mathcal{M}$, which is such that the strings pierce an embedded axionic domain wall across which the $\theta$ parameter changes.\footnote{Similar setups are also realized by junctions of matter in condensed matter systems (see e.g. \cite{josephson}) and in non-Abelian cases (see e.g. \cite{Dierigl:2014xta}).} Second, we study a limiting case of certain superconducting Abelian cosmic strings in the bulk, again with both endpoints ending on the boundary, but in this case with no embedded domain wall present. Throughout, we work in the limit in which backreaction of the strings on the geometry is absent.\footnote{We should note that throughout the paper, we will have in mind a `radial' foliation of the bulk manifold $\mathcal{M}$. This is commonplace in the examples of AdS/CFT duality, which will be a natural example in our mind.}
\newline

Extended objects which obey fractional statistics in theories with topological terms in the Lagrangian have in fact already been studied in the literature in string theory setups (see e.g. \cite{Hartnoll:2006zb}). Although related to our investigations, these studies are different from our discussions, in particular because we are considering a pure quantum field theory setup in $(3+1)$ spacetime dimensions. In contrast, the discussions in \cite{Hartnoll:2006zb} are in the context of $(4+1)$ dimensions in a string theory setup.
\newline

This work is organized as follows. In section \ref{sec:3drev} we study electrically charged $(2+1)$ dimensional Abelian vortices in flat spacetime from two different perspectives. In particular, in subsection \ref{subsec:3dknown} we briefly review the result of \cite{Frohlich:1988qh} that $(2+1)$ dimensional electrically charged Chern-Simons vortices obey fractional statistics as a consequence of being electrically charged. Although well known, we rewrite the steps in the dual field's language, which also sets up our tactics for higher dimensions. In subsection \ref{subsec:3daddcurrent} we argue that $(2+1)$ dimensional vortices with no pure Chern-Simons term can also obey fractional statistics in the presence of an additional internal current. Although straightforward, to our knowledge, this way of looking at charged vortices did not appear in the literature. In section \ref{sec:4d} we then study some particular $(3+1)$ dimensional Abelian cosmic strings in the bulk which end on its boundary (modulo the assumptions discussed in previous paragraphs). We will again consider two ways of charging such string-vortices. In subsection \ref{subsec:4dtheta}, in complete analogy to the $(2+1)$ dimensional case of subsection \ref{subsec:3dknown}, we consider $(3+1)$ dimensional cosmic strings from the dual point of view and show that cosmic strings obey fractional statistics if they are electrically charged. Later on, in subsection \ref{subsec:4daddgauge} we study the above system using the alternate description of an additional gauge field along the strings. This turns out to be closely related to superconducting strings considered earlier by Witten \cite{Witten:1984eb}. We comment on that in subsection \ref{subsec:wittstring}. In section \ref{sec:4d3drel}, we combine our discussions of the above two sections and point out that in the context of a four dimensional bulk, the boundary endpoints of the above-mentioned cosmic strings are nothing but the vortices which we considered in section \ref{sec:3drev}. This is clear from both the perspectives that we mentioned so far, and we discuss them in subsections \ref{subsec:4d3ddual} and \ref{subsec:4d3daddgauge} respectively. We conclude in section \ref{sec:concl}, where we conjecture a general condition which has to be satisfied for $(2+1)$ dimensional vortex endpoints of any $(3+1)$ dimensional cosmic string, in order for the string to obey fractional statistics.

\section{Fractional statistics of Abelian vortices in $(2+1)$ dimensions}\label{sec:3drev}

Equipping vortices with an electric charge generally leads to a contribution to the effective action $\mathcal{S}$ of the Hopf form \cite{wenzee}
\begin{equation}
\label{eq:hopf}
\mathcal{S} \supseteq \mathcal{S}_{Hopf} = \beta \int j_{(vortex) \mu} \frac{\epsilon^{\mu \nu \lambda} \partial_\nu}{\partial^2} J_{(el) \lambda}\,,
\end{equation} 
obtained after integrating out any mediating gauge fields after appropriate gauge fixing. Here, $\beta$ is a theory dependent coefficient.
The Hopf term describes the interaction of a (topological) vortex current $j_{(vortex) \mu}$ whose charge corresponds to the winding number, with an electric current $J_{(el) \lambda}$, also localized on a vortex. In its presence, 
interchanging two vortices with electric charges $q_e^{(i)}$ and winding numbers $n^{(i)}$ implies an Aharonov-Bohm phase
\begin{equation}
\label{eq:abphase}
\Delta {\mathcal S} = \frac{\beta}{2} \left(n^{(1)} q_e^{(2)} + n^{(2)} q_e^{(1)}\right)\,.
\end{equation}
For suitably chosen $\beta$, this equation entails the existence of anyonic vortices.
\newline

In the following, we discuss two explicit realizations of this setup, which are in part well known results (see e.g. \cite{Frohlich:1988qh}). In the first scenario, vortices are equipped with electric charge by means of a Chern-Simons term in their action. In that case, the electric current above is identical to the vortex current. In another setup, the vortex is explicitly charged under an additional gauge field.

\subsection{Electrically charged Chern-Simons vortices}\label{subsec:3dknown}

Electrically charged static Abelian vortices in $(2+1)$ dimensional Minkowski spacetime can be obtained numerically as static lowest energy topologically non-trivial configurations in the theory with a Chern-Simons term added to the Nielsen-Olesen Lagrangian \cite{Paul:1986ix}:
\begin{equation}
\mathcal{L} = - \frac{1}{4} F_{\mu \nu}F^{\mu \nu} + \frac{1}{2} \left(D_\mu \phi \right)^\dagger \left(D^\mu \phi \right) - \frac{\lambda}{4} \left(\phi^\dagger \phi - v^2\right)^2 + \mu \epsilon^{\mu \nu \alpha} A_\mu \partial_\nu A_\alpha \, .
\label{lagrangiancs}
\end{equation}
Here $F_{\mu \nu} \equiv \partial_{[\mu} A_{\nu]}$, $D_\mu  \phi \equiv \partial_\mu \phi - i e A_\mu \phi$ and $\phi$ is a complex scalar field which can be parametrized as
\begin{equation}
\phi(x) = \rho(x) e^{i \theta_{(v)}(x)} \, ,
\end{equation}
with the two real valued functions $\rho(x)$ and $\theta_{(v)}(x)$. This Lagrangian is invariant under the $U(1)$ transformation $\phi \rightarrow e^{iw(x)} \phi$, $A_\mu \rightarrow A_\mu + \frac{1}{e}\partial_\mu w(x)$ with $w(x)$ the gauge parameter.
\newline

Electrically charged vortices with the electric current $J_\mu$ given by
\begin{equation}
J_\mu = \frac{ie}{2} \left(\phi \left(D_\mu \phi\right)^\dagger - \phi^\dagger D_\mu \phi\right) \, ,
\label{currentor}
\end{equation}
have been found in \cite{Paul:1986ix} numerically by minimizing the energy functional which corresponds to this Lagrangian after using appropriate ansatz-functions and boundary conditions for the scalar and gauge field. The electric charge is 
\begin{equation}
Q_e \equiv \int J_0 \,d^2x = 2 \mu \Phi_B \, ,
\label{elc}
\end{equation}
where $\Phi_B \equiv \int d^2 x \epsilon^{ij} \partial_i A_j$ is the magnetic flux. The vortices carry finite energy since the solutions for $A_\mu$, $\rho$ and $\theta_{(v)}$ are such that at spatial infinity $\rho(x) \rightarrow v$ and $\left(\partial_\mu \theta_{(v)} - e A_\mu\right) \rightarrow 0$ \cite{Paul:1986ix}.%
\newline

For simplicity, we can approximate the vortex to be point-like, assuming that $\rho(x) \equiv v$ everywhere except at $x = 0$. More general treatments can be found for example in \cite{Kim:1992yz} and in references therein. In this approximation the Lagrangian (\ref{lagrangiancs}) can then be written as
\begin{equation}
\mathcal{L} = \frac{v^2}{2} \left(\partial_\mu \theta_{(v)} - e A_\mu \right)^2 - \frac{1}{4} F_{\mu \nu}F^{\mu \nu} + \mu \epsilon^{\mu \nu \alpha} A_\mu \partial_\nu A_\alpha \, .
\label{lag}
\end{equation}
The electric current (\ref{currentor}) becomes
\begin{equation}
J_\mu = e v^2 \left(\partial_\mu \theta_{(v)} - e A_\mu \right) \, .
\label{current}
\end{equation}

At low energies, one can dualize the Lagrangian (\ref{lag}) and in the dual theory the vortices appear as point charges of a gauge field $B_\mu$. In fact, going to the dual picture is very useful to also visualize that (if the constant $\mu$ in (\ref{lag}) is appropriately chosen) electrically charged vortices obey fractional statistics \cite{Wen:1988uw}, a result which is well known \cite{Frohlich:1988qh} and which could also be inferred directly from (\ref{lag}).
\newline

The typical way to dualize (\ref{lag}) is to introduce an auxiliary field $J^{(aux)}_\mu$ and rewrite (\ref{lag}) as
\begin{equation}\label{eq:aux}
\mathcal{L} = - \frac{1}{2v^2} J^{(aux)}_\mu J^{(aux)\mu} + J^{(aux)}_\mu \left(\partial^\mu \theta_{(v)} - e A^\mu\right) + \mu \epsilon^{\mu \nu \alpha} A_\mu \partial_\nu A_\alpha \,.
\end{equation} 
If we introduce a dual $U(1)$ gauge field $B_\mu$ like\footnote{The equations of motion for the auxiliary field give $J^{(aux)}_\mu = \frac{1}{e} J_\mu$. Therefore, using Stokes theorem, the electric charge can be written as
\begin{equation}
Q_e = 2 \mu \Phi_B = \int J_0 d^2 x = \frac{e}{2\pi}\int \partial_iB_j \epsilon^{ij} d^2x = \frac{e}{2\pi}\oint B_\mu dx^\mu \, .
\label{chargee}
\end{equation}} 
\begin{equation}
J^{(aux)}_\mu = \frac{1}{2\pi}\epsilon_{\mu \nu \alpha} \partial^\nu B^\alpha \, ,
\end{equation}
and retain only the topological contributions relevant at low energies,
the Lagrangian becomes
\begin{equation}
\mathcal{L} = 
 - \frac{1}{2\pi}dB \wedge \left(d\theta_{(v)} - e A\right) + \mu A \wedge dA \, ,
\label{lagbeforeintegrating}
\end{equation}
where we have now adopted the usual index free notation for differential forms to simplify expressions.
Integrating out $A$, we obtain
\begin{align}
\mathcal{L} &= 
-\frac{1}{2\pi}dB\wedge d\theta_{(v)} + \tilde{\mu}B\wedge dB \nonumber \\
&=  B\wedge\star j_{(vortex)} + \tilde{\mu} B \wedge dB \, ,
\label{duallag}
\end{align}
with $\tilde{\mu} \equiv -\frac{e^2}{16\pi^2 \mu}$. In the second line, we have integrated by parts 
and introduced the vortex current
\begin{equation}
j_{(vortex)} \equiv \frac{1}{2\pi} \,{\star dd\theta_{(v)}}\,,
\end{equation}
which is conserved and non-zero for a vortex since $\theta_{(v)}$ is not single valued.\footnote{Consequently, $\theta_{(v)}$ is not a zero-form and thus not annihilated by $dd$. In other words, since $\theta_{(v)}$ is discontinuous, partial derivatives do not commute everywhere. This leads to a non-trivial current localized on the vortex.} Its associated charge $\int d^2x\,j^0_{(vortex)}$ is the winding number of the vortex configuration.
\newline

We observe that the Chern-Simons term manifests itself also in the dual theory at low energies, a well known result sometimes referred to as ``Chern-Simons self-duality". 
Moreover, the vortex current appears as an ``electric" current for the gauge potential $B$.
Note that both the interaction term and the Chern-Simons term contribute to the statistical phase induced in an exchange process of two identical currents \cite{Kim:1992yz, Wen:1988uw, wenzee}.
\newline

The equations of motion for the dual gauge field $B_\mu$ are
\begin{equation}
j_{(vortex)} =  2\tilde{\mu}\, {\star dB} \, .
\label{eomm}
\end{equation}
Integrating out $B$ using a suitably gauge fixed version of the above directly yields \eqref{eq:hopf} with $\beta = -\frac{1}{4\tilde{\mu}}$ and $J_{el} = j_{(vortex)}$. Consequently, $q^{(i)}_e = n^{(i)}$ in \eqref{eq:abphase} and interchanging two identical vortices with winding number $n^{(i)} = 1$ yields an Aharonov-Bohm phase 
\begin{equation}
\Delta {\mathcal S} = -\frac{1}{4\tilde{\mu}}\,.
\end{equation}
Note that we can relate this phase to the electric charge $Q_e$ as defined in \eqref{elc} via
\begin{equation}
Q_e = \frac{1}{2\pi}\int d^2x \epsilon^{ij} \partial_iB_j   = -\frac{1}{4\pi\tilde{\mu}}  \int d^2 x j_{(vortex) 0}  = -\frac{e}{4\pi\tilde{\mu}}  \,,
\end{equation}
to give
\begin{equation}
\label{eq:phase_ec}
\Delta {\mathcal S} =\frac{\pi}{e} Q_e \,.
\end{equation}

Alternatively, the statistical Aharonov-Bohm phase induced in the vortex exchange process can also be extracted via the introduction of a ``total current" $j^{tot}$ as \cite{wenzee}
\begin{equation}
j^{tot} \equiv  j_{(vortex)} + \tilde{\mu}\,{\star dB} \, .
\label{totalcurrent}
\end{equation}
This total current is the coefficient of $d\Lambda(x)$ in the variation of (\ref{duallag}) under $B \rightarrow B + d\Lambda$. The total charge $Q \equiv  \int d^2x j_0^{tot}$, is then given by (using (\ref{eomm}))
\begin{equation}
Q = \frac{1}{2} \int d^2 x j_{(vortex) 0} = \frac{n}{2}\, ,
\label{totalcharge}
\end{equation}
with the winding number $n$.
If one vortex $(1)$ with charge $Q = \frac{1}{2}$ is taken once around another identical vortex (denoted by superscript $(2)$) at rest which produces the potential $B^{(2)}$, an Aharonov-Bohm phase
\begin{equation}
e^{i \frac{1}{2}\oint B^{(2)}} = e^{i \frac{\pi}{e} \int d^2x J_0^{(2)}} = e^{i \pi \frac{2\mu}{e} \Phi_B}
\label{eqq}
\end{equation}
is induced.
Here the line integral is taken along the contour (worldline) of vortex $(1)$. 
\newline

Including an additional factor of $2$ to account for the potential generated by vortex $(1)$, we obtain for the phase of the exchange process
\begin{equation}
\Delta \mathcal{S} = \pi \frac{2\mu}{e} \Phi_B = \frac{\pi}{e} Q_e\,,
\label{chac}
\end{equation}
in agreement with \eqref{eq:phase_ec}.
Thus, in that case, the vortices obey fractional statistics (note that if the Chern-Simons term in (\ref{lagrangiancs}) is absent, i.e. if $\mu = 0$ in (\ref{lagrangiancs}), then $\Delta \mathcal{S} = 0$).
\newline

Let us emphasize that the use of the total current (\ref{totalcurrent}) makes sure that both the phase shifts generated by the interaction $j_{(vortex) \mu}B^\mu$ and by the Chern-Simons term are taken into account \cite{Goldhaber:1988iw}. In our setup, the use of the total current instead of the vortex current produces only an additional factor of $\frac{1}{2}$ \eqref{totalcharge}. In other similar setups, not using the total current can however lead to qualitatively wrong conclusions, as discussed in  \cite{Goldhaber:1988iw}.
\newline

We end this section by pointing out that the result could equivalently have been obtained by integrating out $B$ from  \eqref{lagbeforeintegrating} using the equation of motion for $A$. The resulting action for $A$ then (of course) takes on the exact same form \eqref{duallag}. This will prove useful in later sections.

\subsection{Charging Abelian vortices using additional current}\label{subsec:3daddcurrent}

Vortices in $(2+1)$ spacetime dimensions which obey fractional statistics can also be obtained in a different way, without the presence of a pure Chern-Simons term. One obvious 
example is the case of a mixed Chern-Simons term in the dual theory with an additional gauge field $E_\mu$ and an internal current $\tilde{J}_\mu$ which is localized on the vortex and coupled to $E_\mu$. The corresponding dual low energy Lagrangian (without kinetic terms) is given by\footnote{This setup can also be generalized by adding pure Chern-Simons terms for $B_\mu$ and/or for $E_\mu$ to (\ref{laggran}). The statistical phase which we will determine then changes accordingly. }
\begin{equation}
\mathcal{L} = B \wedge\star j_{(vortex)} + E\wedge\star\tilde{J} + \kappa E\wedge dB \, .
\label{laggran}
\end{equation}

Integrating out both gauge fields directly gives rise to \eqref{eq:hopf} with $\beta = -1/\kappa$. Consequently, interchanging two identical vortices with electric charge $q_e = \int \tilde{J}_0 d^2 x$ and winding number $n$ gives rise to the statistical phase
\begin{equation}
\label{eq:statphaseel}
\Delta {\mathcal S} = -\frac{1}{\kappa} n q_e\,.
\end{equation}

Alternatively, as before, the statistical phase due to the interchange of two vortices can be expressed as
\begin{equation}
\Delta \mathcal{S} = \oint B^{(2)}\, ,
\label{shift}
\end{equation}
where the line integral is once again taken along the worldline element of the vortex $(1)$ and $B^{(2)}$ is the potential sourced by vortex (2). 
Here, a factor of $\frac{1}{2}$ due to considering an exchange process was canceled by a factor of $2$ due to the contribution to the potential sourced by string $(1)$.
Using the equations of motion for $E_\mu^{(2)}$,
\begin{equation}
\tilde{J} = - \kappa\, {\star dB} \, ,
\label{eomJ}
\end{equation}
and applying Stokes theorem,
the line integral yields
\begin{equation}
\Delta \mathcal{S} = -\frac{1}{\kappa} q_e \, ,
\label{dlll}
\end{equation}
in agreement with \eqref{eq:statphaseel}.
Therefore, if the charge $\int d^2 x \tilde{J}_0^{(2)}$ of the vortex at rest is non-vanishing and if $\kappa$ is chosen appropriately such that a non-trivial Aharonov-Bohm phase shift is induced by $\Delta \mathcal{S}$, these vortices obey fractional statistics.
\newline

There are microscopic models that give rise to an effective Lagrangian of the form \eqref{laggran}. In particular, it
appears naturally in a boundary viewpoint of certain superconducting cosmic strings \cite{Witten:1984eb}, as we will point out in section \ref{sec:4d3drel}.

\section{Fractional statistics of string-like vortices in the bulk}\label{sec:4d}

In $(3+1)$ dimensional flat spacetime cosmic strings exist as static topologically non-trivial lowest energy configurations in the theory given by the Nielsen-Olesen Lagrangian \cite{Nielsen:1973cs}. Such Nielsen-Olesen cosmic strings have also been studied as solutions in other spacetimes, e.g. in global AdS, both with and without the backreaction of the cosmic string on the spacetime taken into account \cite{Dehghani:2001ft}.
\newline

A dualization argument analogous to the one presented in section \ref{sec:3drev} goes through for cosmic strings in $(3+1)$ spacetime dimensions \cite{Franz:2006gb}: let us consider a cosmic string in flat spacetime with electric charge
\begin{equation}
\int J_0 d^3 x = e v^2 \int \left(\partial_0 \theta_{(v)} - e A_0\right) d^3 x \, .
\end{equation}
Once again, we can introduce an auxiliary field $J^{(aux)}_\mu$ which can be written as
\begin{equation}
J^{(aux)}_\mu = \frac{1}{2 \pi} \epsilon_{\mu \nu \alpha \beta} \partial^\nu B^{\alpha \beta}
\end{equation}
for some two-form $B_{\mu \nu}$. In complete analogy to the case of vortices in $(2+1)$ spacetime dimensions which we have considered in section \ref{sec:3drev}, $J^{(aux)}_\mu$ can be identified with the electric current $J_\mu$ of the cosmic string and its electric charge can thus, in analogy to (\ref{chargee}), be written in terms of $B_{\mu \nu}$ as
\begin{equation}
\int J_0 d^3x = e \int \partial_i B_{jk} \epsilon^{ijk} d^3x = e \oint B_{ij} dx^i \wedge dx^j \, .
\label{cur}
\end{equation}
As for $(2+1)$ dimensional pure Nielsen-Olesen vortices, the electric charge for pure Nielsen-Olesen cosmic strings is zero \cite{Julia:1975ff}.
\newline

Let us now consider two particular models for cosmic strings (different from the Nielsen-Olesen type) in $\mathcal{M}$ which are electrically charged, and by dualizing their corresponding Lagrangians show that in these cases the strings obey fractional statistics. As we will discuss in section \ref{sec:4d3drel}, these electrically charged cosmic strings, in contrast to the pure Nielsen-Olesen cases, have boundary vortices/antivortices as endpoints which carry different electric charges.
\newline

\subsection{$U(1)$ charged string-vortices with $\theta$ terms}\label{subsec:4dtheta}

In this subsection we will first consider a theory with Nielsen-Olesen Lagrangian with an added topological term $\Delta \mathcal{L} = \theta \epsilon^{\mu \nu \alpha \beta} F_{\mu \nu} F_{\alpha \beta} = \theta F \wedge F$ in $\mathcal{M}_4$ (with $g$ the metric and $\epsilon^{\mu \nu \alpha \beta}$ the Levi-Civita symbol)\footnote{Throughout this section we shall use $\epsilon^{\mu \nu \alpha \beta}$ for the Levi-Civita symbol and not for the Levi-Civita tensor. Also note that, as mentioned before, throughout this paper we will neglect the backreaction on the metric and not consider any Einstein-Hilbert term or associated boundary counterterms.},
\begin{equation}
\mathcal{L} = \sqrt{-g}\left(\frac{1}{4} F\wedge \star F + \frac{1}{2} \left(D_\mu \phi \right)^\dagger \left(D^\mu \phi \right) - \frac{\lambda}{4} \left(\phi^\dagger \phi - v^2\right)^2\right) + \theta F \wedge F \, .
\label{thetalag}
\end{equation}

\textbf{Constant $\theta$}: If $\theta$ in (\ref{thetalag}) is a \emph{constant} parameter, the term $\theta F \wedge F$ is a pure boundary term,
\begin{equation}
\theta F \wedge F = d \left(\theta  A \wedge F \right) \, .
\label{boundarycs0}
\end{equation}
It thus has no effect on the bulk equations of motion \cite{Witten:2003ya} and the standard Nielsen-Olesen cosmic strings
are topologically non-trivial lowest energy configurations. Such pure Nielsen-Olesen cosmic strings do not obey fractional statistics as a consequence of not being electrically charged \cite{Julia:1975ff}. We shall show in section \ref{sec:4d3drel} that the absence of fractional statistics in this case is consistent with the statistics of the induced boundary theory. In other words, the Nielsen-Olesen cosmic string solution of (\ref{thetalag}) with constant $\theta$ amounts to boundary endpoint vortices which \emph{do not obey} fractional statistics.
\newline

\textbf{Non-constant $\theta$}: However, in the case of a \emph{non-constant} $\theta$, it is not a pure boundary term, but contains an extra contribution:
\begin{equation}
\theta F \wedge F = d \left(\theta  A \wedge F \right) - d\theta \wedge A \wedge F \, .
\label{boundarycs}
\end{equation}
Let us consider a cosmic string which is oriented in the $z$-direction and ends on both sides on some $(2+1)$ dimensional boundary $\partial\mathcal{M}_4$. For simplicity, we take the parameter $\theta$ to be constant everywhere except at one axionic domain wall embedded in the bulk along which $\theta$ changes from a value $\theta_u$ to another value $\theta_d$.\footnote{Note that in such a setup the first term on the right hand side of (\ref{boundarycs}) induces a Chern-Simons term on the spacetime boundary \cite{Witten:2003ya} whereas the second term on the right hand side of (\ref{boundarycs}) induces a Chern-Simons term on the domain wall \cite{Dierigl:2014xta}.} In this case, the Nielsen-Olesen cosmic string gets electrically charged in a manner similar to the Witten effect (but applied to $U(1)$) which says that a $\theta$ term in the Lagrangian can induce electric charges for magnetic monopoles \cite{Witten:1979ey} and also for certain other topological solitons \cite{josephson}. The electric charge density $\rho$ induced by the term $\Delta \mathcal{L}$ can be directly read off from the corresponding Maxwell equation:
\begin{equation}
\rho = 
d\theta \wedge dA = B_{mag}\, \partial_z \theta \, ,
\end{equation}
where ${B}_{mag}$ is the amplitude of the divergence-free magnetic field of the cosmic string.
The electric charge $Q_E$ is  defined through the integral of $\rho$ on an arbitrary Cauchy surface $\Sigma$. Given that $\rho$ is a total derivative, this can be written as
\begin{equation}
Q_E \equiv \int_\Sigma \rho  = \int_{\partial\Sigma} \theta B_{mag}  = \Phi_B \left(\theta_u - \theta_d\right) \, ,
\label{electric}
\end{equation}
with $\Phi_B$ the magnetic flux through the string and $\theta_u$ and $\theta_d$ the values of $\theta$ at the upper and lower endpoints of the cosmic string. 
We observe that the electric charge of the string is a boundary term. This feature will allow us to argue that the statistical properties of the string are completely determined by boundary physics. In order to see this more explicitly, we begin by demonstrating using the dual picture that 
electrically charged cosmic strings 
	obey
 fractional statistics.
\newline

Dualizing the Lagrangian (\ref{thetalag}) at low energies (i.e. considering only the topological terms) in the approximation analogous to the one used in section \ref{subsec:3dknown} and using the auxiliary field $J^{(aux)}_\mu \equiv \frac{1}{2\pi\sqrt{-g}}\epsilon_{\mu \nu \alpha \beta} \partial^\nu B^{\alpha \beta}$, we get
\begin{equation}
\mathcal{L} = 
- \frac{1}{2\pi} dB\wedge \left(d\theta_{(v)} - e A\right) + \theta\, dA \wedge dA \, .
\label{laggg}
\end{equation}
We can now integrate out $B$ using the equation of motion for $A$, as hinted already at the end of section \ref{subsec:3dknown}. We obtain
\begin{equation}
dB = \frac{4\pi}{e} d\theta\wedge dA\,,
\end{equation}
and consequently
\begin{equation}
\mathcal{L} = 
-\frac{2}{e} d\theta\wedge dA\wedge \left(d\theta_{(v)} - e A\right) + \theta dA \wedge dA \, .
\label{laggg2}
\end{equation}
Integrating by parts and considering only the topological contributions let us rewrite this as
\begin{equation}
\mathcal{L} = \mathcal{L}_{bulk} + \mathcal{L}_{bdy} \, ,
\end{equation}
with
\begin{equation}
\mathcal{L}_{bulk} \equiv \frac{4\pi}{e}\theta \,dA\wedge \star j_{(vortex)} + d\theta\wedge A\wedge dA \, ,
\label{bulke}
\end{equation}
and
\begin{equation}
\mathcal{L}_{bdy} \equiv d\left(-\frac{2}{e}\,\theta\, dA \wedge d\theta_{(v)} + \theta A\wedge dA\right) \, .
\label{bo}
\end{equation}
Here $j_{(vortex)} \equiv \frac{1}{2\pi}\star dd\theta_{(v)}$ can be interpreted as a vortex loop current. The normalizing factor was chosen as to guarantee that the associated charge is integer valued. We can see that $\theta$ not being constant allows a coupling of the two-form loop current $j_{(vortex)}$ to the one-form $A$, for which for the same reason a Chern-Simons-like term can be constructed. In other setups, non-trivial statistics for strings would require a coupling to a two-form $B$, for which a Chern-Simons term can be written only in higher dimensions.

What we called ``$\mathcal{L}_{bulk}$" is equivalent to the whole Lagrangian $\mathcal{L}$ of the theory in the case of a manifold without boundary. What we called ``$\mathcal{L}_{bdy}$" are the additional contributions which have to be taken into account due to the presence of the boundary such that the whole Lagrangian $\mathcal{L}$ in this case of a manifold with boundary is given by $\mathcal{L} = \mathcal{L}_{bulk} + \mathcal{L}_{bdy}$. We shall use analogous definitions for $\mathcal{L}_{bulk}$ and $\mathcal{L}_{bdy}$ in later sections. For the rest of this section, we focus on $\mathcal{L}_{bulk}$ and we will come back to the contributions of $\mathcal{L}_{bdy}$ in section \ref{sec:4d3drel}.
\newline

Note that we can also integrate out $A$ in $\mathcal{L}_{bulk}$ to obtain the effective Lagrangian
\begin{equation}
\label{eq:bulkano}
\mathcal{L}_{bulk} = -\frac{4 \pi}{e^2} d\theta\wedge \star j^{(vortex)} \wedge d\theta_{(v)} \, .
\end{equation}
Given the definition of the vortex current, this resembles the Hopf term in lower dimensions if written in coordinates,
\begin{equation}
\mathcal{L}_{bulk} \sim \partial_\mu \theta \epsilon_{\nu\alpha\beta\gamma} j^{(vortex) \mu \nu} \frac{\partial^\alpha}{\Box} j^{(vortex) \beta\gamma} \, .
\end{equation}

In a process in which one cosmic string $(1)$ is adiabatically taken around another identical one $(2)$ at rest\footnote{As we have mentioned in the introduction, in our context such processes are meaningful because, by design, one of the string is static and the other one is extended radially inward from the boundary without any intersections between their worldsheets.} in such a way that the positions of the strings are exchanged, \eqref{eq:bulkano} induces a change in the action of
\begin{equation}
\label{eq:bulkph}
\Delta \mathcal{S} = \frac{4 \pi}{e^2}\oint d\theta \wedge d \theta_{(v)}^{(2)} = \frac{4 \pi}{e} \int d \theta \wedge d A^{(2)} = \frac{4 \pi}{e} Q_E \, ,
\end{equation}
where the surface integral is
 localized on the trajectory of string $(1)$ and the volume integral, obtained by Stokes' theorem, to its interior. Moreover, $\theta^{(2)}_{(v)}$ and $A^{(2)}$ are the $\theta_{(v)}$ parameter and gauge potential of string $(2)$. Again, the contribution from $A^{(1)}$ of string $(1)$ has canceled a factor of $1/2$. The last two equalities follow from the equation of motion for $B$ which can be obtained from equations (\ref{laggg}) and \eqref{electric}. 
\newline

\subsection{Charging $U(1)$ string-vortices using additional current}\label{subsec:4daddgauge}

In complete analogy to the $(2+1)$ dimensional vortices of subsection \ref{subsec:3daddcurrent}, cosmic strings in $(3+1)$ spacetime dimensions can also be 
endowed with fractional statistics by coupling it two a two-form $B$ and adding a one-form gauge field $E$, as well as an internal one-form current $\tilde{J}$ localized on the string to which $E$ couples. A mixed Chern-Simons term between $E$ and $B$ can be constructed and gives rise to fractional statistics.
This setup has been studied in the past for flat spacetime \cite{Aneziris:1990gm}.
\newline

The corresponding dual low energy bulk Lagrangian is given by\footnote{Just as in subsection \ref{subsec:3daddcurrent}, the present setup can also be generalized. For example, one can add a term $\epsilon^{\mu \nu \alpha \beta} B_{\mu \nu} B_{\alpha \beta}$ and/or a term $\epsilon^{\mu \nu \alpha \beta} \partial_\mu E_\nu \partial_\alpha E_\beta$ to (\ref{lagsup}). The statistical phase  then changes accordingly.}
\begin{equation}
\mathcal{L} = B \wedge \star j_{(vortex)} +  E \wedge \star\tilde{J} + \theta dE\wedge B \, .
\label{lagsup}
\end{equation}
This Lagrangian is the higher dimensional analogue of \eqref{laggran}. 
As shown in \cite{Aneziris:1990gm}, it describes
cosmic strings that obey fractional statistics if the constant $\theta$ and the current $\tilde{J}_\mu$ in (\ref{lagsup}) are such that the charge $\int \tilde{J}_0$ is not an integer-multiple of $2 \pi \theta$. To see this, we note that the equations of motion for $E_\mu$ are given by
\begin{equation}
\tilde{J} = - \theta\, {\star dB}  \, .
\end{equation}
Therefore, in a process in which one cosmic string is adiabatically taken around another identical one at rest such that the initial positions of the strings get exchanged, the interaction term $B\wedge\star j_{(vortex)}$ induces a change in the action of the form
\begin{equation}
\Delta \mathcal{S} =  \oint B = -\frac{1}{\theta} \int d^3 x \tilde{J}_0 \, ,
\label{deltt}
\end{equation}
where Stokes theorem was used. $B$ is the field corresponding to the string at rest. The surface integral is taken along the worldsheet of the moving string. 

In the presence of a boundary, the Lagrangian \eqref{lagsup} cannot be complete, as it is not gauge invariant. Instead, it changes by a boundary term. On similar grounds, the charge corresponding to $\tilde{J}$ is not conserved since it can be exchanged with the boundary. Correspondingly, we need to supply the theory with boundary contributions of the form
\begin{equation}
{\cal L}_{bdy} = B \wedge \star j_{(vortex)} +  E \wedge \star\tilde{J}_{bdy} + \theta dE\wedge B \,,
\end{equation}
where now $B$ is a boundary one-form that shifts under the gauge transformation of the bulk two-form as $B \to B - \Lambda$, $j_{(vortex)}$ is the boundary vortex current and $E$ is obtained by taking the bulk one-form $E$ to the boundary. The current $\tilde{J}_{bdy}$ is defined through the requirement that the total charge $Q_{tot} \equiv \int_\Sigma \tilde{J} + \int_{\partial\Sigma} \tilde{J}_{bdy}$ is constant. Obviously, $\tilde{J}_{bdy}$ is not unique and can be shifted by an arbitrary divergence free vector.
\newline

We observe that the boundary Lagrangian bears close resemblance with \eqref{laggran}. Moreover, by an appropriate shift of $\tilde{J}_{bdy}$, it is always possible to choose the latter such that $Q_{tot} = 0$. It is in this case that the phase shift of the boundary precisely matches that of the bulk theory upon appropriate identification of
 the above $\theta$ with the Chern-Simons coefficient $\kappa$ (\ref{laggran}).
We will elaborate on this relation between the bulk and boundary phase shifts in section \ref{sec:4d3drel}. 
\newline

In the next subsection, we shall demonstrate that the strings in this setup can be related to the superconducting cosmic strings which were introduced in \cite{Witten:1984eb} in spacetimes in which both the strings considered in \ref{subsec:4daddgauge} and the superconducting cosmic strings exist as solutions.

\subsection{Abelian superfluid cosmic string}\label{subsec:wittstring}

In this subsection, we present a microscopic realization of the electrically charged cosmic strings discussed in Sec.~\ref{subsec:4daddgauge}. The model is a slight modification of the superconducting cosmic strings with Bose charge carriers of \cite{Witten:1984eb} (see also \cite{Balachandran:1979cb}). The presentation is divided into two parts. First, we consider a simplified version of \cite{Witten:1984eb}, which equips cosmic string with an additional global charge. We then discuss how this global charge can be coupled to the magnetic flux of the string in order to give rise to fractional statistics.
We restrict the discussion in this subsection to spacetimes in which such superconducting cosmic strings exist and for notational simplicity we shall denote the metric compatible derivative by $\partial_\mu$.
\newline

In order to localize an additional current on the string solutions of the Nielsen-Olesen Lagrangian in $(3+1)$ spacetime dimensions, one introduces a second complex scalar field $\psi$ with appropriate potentials, chosen as to guarantee condensation of this additional scalar on the string.
The corresponding Lagrangian reads
\begin{equation}
\mathcal{L} = \sqrt{-g}\left(- \frac{1}{4} F_{\mu \nu}F^{\mu \nu} + \frac{1}{2}|\left(\partial_\mu - ie A_\mu \right) \phi|^2  + \frac{1}{2}| \partial_\mu  \psi|^2 - V(\phi, \psi)\right)\,.
\label{superstring}
\end{equation}
The potential is parametrized as \cite{Witten:1984eb}
\begin{equation}
V(\phi, \psi) = \lambda_\phi \left(|\phi|^2 - v_\phi^2\right)^2 + \lambda_\psi \left(|\psi|^2 - v_\psi^2\right)^2 + \kappa |\psi|^2|\phi|^2 \, ,
\label{pote}
\end{equation}
and the constants $\kappa$, $\lambda_\psi$, $\lambda_\phi$, $v_\phi$ and $v_\psi$ are chosen such that
\begin{equation}
\lambda_\phi v_\phi^4 > \lambda_\psi v_\psi^4, \quad\frac{2 \kappa}{\lambda_\psi} > \frac{v_\psi^2}{v_\phi^2} \, .
\label{eq}
\end{equation}
The Lagrangian (\ref{superstring}) is invariant under the gauge transformation
\begin{equation}
A_\mu \longrightarrow A_\mu + \frac{1}{e} \partial_\mu \Lambda(x),\quad \phi \longrightarrow e^{i \Lambda(x)} \phi
\end{equation}
as well as the global $U(1)$ rotation
\begin{equation}
\psi \longrightarrow e^{i\varphi} \psi \, .
\end{equation}

The potential (\ref{pote}) and its parameters were chosen such that the $U(1)$ symmetry $A_\mu$ is Higgsed
 (since $\lambda_\phi v_\phi^4 > \lambda_\psi v_\psi^4$ ),
but that the global $U(1)$ of $\psi$ remains unbroken 
(since $\frac{2 \kappa}{\lambda_\psi} > \frac{v_\psi^2}{v_\phi^2}$). In other words, in the given parameter domain, the potential (\ref{pote}) has a minimum at $\phi = v_\phi$, $\psi = 0$.
\newline

It has been demonstrated in \cite{Witten:1984eb} that 
this setup admits stable minimal energy configurations such that $\phi$ and $A_\mu$ describes a cosmic string and $|\psi|$ outside of the string core is (close to) zero, 
but non-zero close to the string core. For a string lying in the z-direction for such configurations, $\psi$ can be parametrized as
\begin{equation}
\psi = |\psi_0(x,y)| e^{i \alpha(z,t)} \, ,
\label{parame}
\end{equation}
where $|\psi_0(x,y)|$ is such that it is exponentially decaying outside of the cosmic string. $|\psi_0(x,y)|$ is a configuration which minimizes the energy of the configuration and $\alpha(z,t)$ parametrizes low-energy excitations which are responsible for making the string superconducting.
\newline

If we use, as in the previous sections, the approximation that the modulus of the scalar field $\phi$ is constant, $\phi = v_\phi$, outside of the string core, one can parametrize it there as
\begin{equation}
\phi = v_\phi e^{i \theta_{(v)}} \, ,
\label{paraphi}
\end{equation}
with the real valued function $\theta_{(v)}$. There are then two conserved currents due to Noether:

\begin{equation}
J_\mu = \sqrt{-g} v_\phi^2 \left(\partial_\mu \theta_{(v)} - e A_\mu\right) \quad\text{and}
\end{equation}
\begin{equation}
\tilde{J}_\mu =  \sqrt{-g} \frac{i}{2}\left(\psi^\dagger\partial_\mu\psi - \psi \partial_\mu \psi^\dagger \right)\, .
\label{ccccc}
\end{equation}
Thus, since $|\psi|$ is unsuppressed only close to the string core, the current $\tilde{J}_\mu$ is effectively localized on the string. Correspondingly, strings in this setup carry an additional global charge.
\newline

While this setup has allowed for an additional global current to be localized on the string, it does not yet imply fractional statistics. 
In order for a nontrivial Aharonov-Bohm phase to emerge, this global current needs to be coupled to the magnetic flux of the string, and thus to the gauge field $A_\mu$. As a proof of concept, let us assume that neither the existence of the string solution and the localization of the additional current is spoiled by a weak minimal coupling of $\psi$ to $A_\mu$. Under this deformation, the Lagrangian \eqref{superstring} changes to
\begin{equation}
\mathcal{L} = \sqrt{-g}\left(- \frac{1}{4} F_{\mu \nu}F^{\mu \nu} + \frac{1}{2}|\left(\partial_\mu - ie A_\mu \right) \phi|^2  + \frac{1}{2}| \left(\partial_\mu - i\tilde{e} A_\mu \right) \psi|^2 - V(\phi, \psi)\right)\,,
\label{superstring_gauged}
\end{equation}
while the current $\tilde{J}_\mu$ becomes on a string solution
\begin{equation}
\tilde{J}_\mu =  \sqrt{-g} \psi_0^2\left(\tilde{e} A_\mu - \partial_\mu\alpha\right) \, .
\label{eq:jmu_string}
\end{equation}

As before, statistical phases become most apparent in a dual picture. We 
dualize the Lagrangian (\ref{superstring})
by introducing a two-form field $B_{\mu \nu}$ and an auxiliary field $J^{(aux)}_\mu$ such that
\begin{equation}
J^{(aux)}_\mu = \frac{1}{\sqrt{-g}} \epsilon_{\mu \nu \alpha \beta} \partial^\nu B^{\alpha \beta}\,.
\end{equation}
Following analogous steps as in the previous sections, one obtains the effective low energy Lagrangian
\begin{equation}
\mathcal{L} = \left(-e A +d\theta_{(v)}\right)\wedge dB +\tilde{e} A \wedge \star\tilde{J} - d\alpha\wedge \star\tilde{J}\, ,
\label{dualsuper}
\end{equation}
where we neglected the kinetic term for $A$
and used 
the explicit form (\ref{paraphi}).
\newline

We now separate the corresponding action into a bulk and a boundary contribution. To this end, we also split the conserved source $\tilde{J}$ into bulk and boundary pieces, $\tilde{J} = \tilde{J}_{bulk} + \tilde{J}_{bdy}$. This split is not unique and the individual contributions need not be conserved; in order to guarantee  the bulk-boundary correspondence of the statistics, additional input is required. It can be made more explicit
by parametrizing
$\tilde{J}$ 
in terms of the string world sheet. To this end, we note that
since $|\psi_0|^2 \approx 0$ outside of the string core, the only significant components of $\tilde{J}_\mu$ are $\tilde{J}_z$ and $\tilde{J}_0$:
\begin{equation}
\tilde{J}_a \equiv \sqrt{-h} \Omega \left(\tilde{e} A_a - \partial_a \alpha\right),\,\, \tilde{J}_x = \tilde{J}_y = 0 \, ,
\end{equation}
with $a \in \{t, z\}$, $\Omega \equiv \int |\psi_0|^2 dx dy$ and $h$ the induced metric on the string. 
This leads to the parametrization \cite{Witten:1984eb, Balachandran:1979cb}
\begin{equation}
\tilde{J}^\mu_{bulk} \equiv \sqrt{-h} \int d^2 \sigma \delta^{(4)}(x-x(\sigma)) \epsilon^{ab}\partial_a x^\mu(\sigma) \partial_b \gamma \, ,
\label{currr}
\end{equation}
with $\sigma^a$ the worldsheet coordinates of the string, $\partial_a \equiv \frac{d}{d \sigma^a}$, $x^\mu$ the string embedding coordinates and $\gamma$ defined via
\begin{equation}
\partial_a \gamma \equiv \Omega \epsilon_{ab}\left(\tilde{e}A^b -\partial^b \alpha\right) \, .
\label{w}
\end{equation}
The corresponding boundary current reads  \cite{Balachandran:1979cb},
\begin{equation}
\label{eq:jbdy}
\tilde{J}^\mu_{bdy} \equiv \int d\sigma^0 \delta^{(4)}(x-x(\sigma^0)) \partial_0 x^\mu(\sigma^0) \gamma(\sigma^0, \sigma^1 = \sigma^1|_{bdy}) \, .
\end{equation}
We note that $\tilde{J}^\mu_{bulk} + \tilde{J}^\mu_{bdy}$ is conserved, as it should be. In addition,
the boundary current \eqref{eq:jbdy} is precisely such that the total charge $Q \equiv \int_\Sigma \tilde{J}_{bulk} + \int_{\partial\Sigma} \tilde{J}_{bdy}$ vanishes. This will prove important when we relate bulk and boundary phase shifts in section \ref{sec:4d3drel}.
\newline

Returning to the action, we obtain
after integration by parts and absorbing a factor of $\tilde{e}$ into $A$
\begin{equation}
\mathcal{S} = \mathcal{S}_{bulk} + \mathcal{S}_{bdy} \, ,
\end{equation}
with
\begin{equation}
\mathcal{S}_{bulk} \equiv \int_\mathcal{M} \left(B\wedge\star j^{(vortex)} + A\wedge{\star\tilde{J}_{bulk}} -\frac{e}{\tilde{e}}\,dA\wedge B\right)
\label{dualsuperbu}
\end{equation}
and
\begin{equation}
\mathcal{S}_{bdy} \equiv \int_{\partial \mathcal{M}}\left(\frac{e}{\tilde{e}} A\wedge B - d\theta_{(v)}\wedge B + A\wedge{\star\tilde{J}_{bdy}} \right)\,,
\label{dualsuperbo}
\end{equation}
where in the former we have made use of the conservation of $\tilde{J}$, i.e. $d{\star\tilde{J}} = 0$, while in the latter we have neglected a term of the form $\alpha\,{\star\tilde{J}}$ that vanishes due to conservation of the total charge.
\newline

Postponing the study of $\mathcal{S}_{bdy}$ to section \ref{sec:4d3drel}, we observe that $\mathcal{S}_{bulk}$ is equivalent to \eqref{lagsup} upon identification of $A$ with $E$ and $\theta$ with $e/{\tilde e}$. In line with the discussion in section \ref{subsec:4daddgauge}, the change
in the action upon interchange of two strings is given by
\begin{equation}
\Delta \mathcal{S} 
= \frac{\tilde{e}}{e} \int d^3 x \tilde{J}_0 \, .
\label{shi}
\end{equation}
Note that 
using the form \eqref{dualsuperbu}, we can write \eqref{shi}
as $\Delta \mathcal{S} \propto \left(\gamma|_{endpoint\,1} - \gamma|_{endpoint \,2}\right)$. 
We will make use of this in section \ref{subsec:4d3daddgauge}.
\newline

Let us end this section on the remark that on the equations of motion for $B$, 
\begin{equation}
\frac{e}{\tilde{e}} dA = \star j^{(vortex)\mu\nu} \, ,
\end{equation}
and for a suitable choice of gauge for $A_\mu$, the action takes on the expected Hopf form
\begin{equation}
\mathcal{L}_{bulk} = \frac{\tilde{e}}{e}\epsilon^{\alpha\beta\mu\nu}j^{(vortex)}_{\alpha\beta}\frac{\partial_\mu}{\Box}\tilde{J}_\nu\,.
\end{equation}

\section{Electrically charged vortices as endpoints of cosmic strings}\label{sec:4d3drel}

In the previous sections we have 
studied the statistics of certain $(2+1)$ dimensional vortices in flat spacetime and of certain $(3+1)$ dimensional cosmic strings in the corresponding bulk. At the end of sections \ref{subsec:4dtheta}, \ref{subsec:4daddgauge} and \ref{subsec:wittstring}, we have already demonstrated how to obtain the boundary statistics of a vortex starting from the bulk statistics in one higher dimension. In this section we will argue that the statistics of the cosmic strings which we found in section \ref{sec:4d}, is exactly the same as the combined statistics of the upper and lower endpoint boundary vortices of the corresponding strings (which we discussed in section \ref{sec:3drev}).
\emph{Thus, the statistics of the cosmic strings in $(3+1)$ dimensional bulk spacetime can be fully understood by considering only the statistics of the boundary vortices of the string on the $(2+1)$ dimensional boundary.} 
\newline

First, we shall consider the case of the cosmic string which can be obtained as classical solution of the Nielsen-Olesen Lagrangian with $\Delta \mathcal{L} = \theta \epsilon^{\mu \nu \alpha \beta} F_{\mu \nu} F_{\alpha \beta}$ added to the Lagrangian (discussed in subsection \ref{subsec:4dtheta}) and the corresponding Chern-Simons vortices (discussed in subsection \ref{subsec:3dknown}). Second, we consider the correspondences between the cosmic strings of subsection \ref{subsec:4daddgauge} and the boundary vortices of subsection \ref{subsec:3daddcurrent}.

\subsection{Bulk string-vortices with $\theta$-term and boundary Chern-Simons vortices}\label{subsec:4d3ddual}

Let us consider the cosmic strings which are obtained as finite energy configurations of the Nielsen-Olesen Lagrangian with an additional term $\Delta \mathcal{L}$.
We will first consider the cosmic strings in the case of a \emph{constant} parameter $\theta$. As mentioned in subsection \ref{subsec:4dtheta}, only for the case of non-constant $\theta$ 
do we obtain non-trivial fractional statistics. Since the boundary term induced by $\Delta\cal{L}$ is an Abelian Chern-Simons term (\ref{boundarycs0}), the endpoints of the cosmic string are nothing but the Abelian Chern-Simons vortices (or antivortices) which we have discussed in subsection \ref{subsec:3dknown}. There we noted that generically, separate Abelian Chern-Simons vortices/antivortices can obey fractional statistics. However, as discussed in section \ref{sec:4d}, the bulk cosmic string does not obey fractional statistics for a \emph{constant} $\theta$. We can reconcile the two by noting that the Aharonov-Bohm phases of the upper and lower endpoint boundary vortices/antivortices always cancel in a process in which one cosmic string is taken around another identical one. 
\newline

To see this clearly and for concreteness, let us consider the two dimensional spatial boundary sphere of conformally compactified AdS$_4$ ($\mathbf{S}^2$) and two cosmic strings in AdS$_4$ which end on this sphere. Let the upper endpoints of the strings end on the northern hemisphere of this 2-sphere and the lower endpoints of the strings end on the southern hemisphere. From the point of view of an observer who is located on this $\mathbf{S}^2$, the upper endpoints are vortices whereas the lower endpoints are antivortices. Since $\theta^\mu_{(vortex)} = - \theta^\mu_{(antivortex)}$, we obtain
\begin{equation}
j^\mu_{(vortex)} = - j^\mu_{(antivortex)} \, .
\end{equation}
Here the $j^\mu$ is once again the vortex current, $j^\mu_{(vortex)} \equiv \epsilon^{\mu \nu \alpha} \partial_\nu \partial_\alpha \theta_{(vortex)}$, which was introduced in (\ref{duallag}).
Using Stokes theorem and the equations of motion $\epsilon^{\mu \nu \alpha}\partial_\nu B_\alpha \propto j^\mu_{(vortex)}$ (\ref{eomm}), the change in the action induced by one upper vortex moving around the other identical one goes as \eqref{eqq}
\begin{equation}
\oint B_\mu dx^\mu_{upper} = \int d^2 x j^0_{(vortex)} \, ,
\end{equation}
whereas the change in the action induced by one lower antivortex moving around the other identical one goes as \eqref{eqq}
\begin{equation}
\oint B_\mu dx^\mu_{lower} = \int d^2 x j^0_{(antivortex)} \, .
\label{sed}
\end{equation}
Here $dx^\mu_{upper}$ is the worldline of a vortex current whereas $dx^\mu_{lower}$ is the worldline of an antivortex current, implying $dx^\mu_{lower} = -dx^\mu_{upper}$.\footnote{In (\ref{sed}) two minus signs cancel: one coming from the change in directions in the curve integration (when compared to the upper case) and other one due to the difference between an antivortex and a vortex: $dx^\mu_{upper} = -dx^\mu_{lower}$.} Since $j^0_{(vortex)} = -j^0_{(antivortex)}$, in total
\begin{equation}
\oint B_\mu dx^\mu_{upper} + \oint B_\mu dx^\mu_{lower} = \int d^2 x \left(j^0_{(vortex)} + j^0_{(antivortex)}\right) = 0 \, .
\end{equation}

Thus, the combined Aharonov-Bohm phase shift of the upper and lower endpoint boundary vortices/antivortices cancel and in this sense the upper and lower boundary endpoint vortices/antivortices of the cosmic string taken together do not obey fractional statistics. Since also the Nielsen-Olesen bulk cosmic strings of the form discussed in subsection \ref{subsec:4dtheta} do not obey fractional statistics for constant $\theta$, the statistics of the boundary endpoint vortices/antivortices matches with the statistics of the bulk cosmic strings in this case.
\newline

Let us now consider the case of a string which ends on both sides on the conformal boundary of AdS$_4$ and is piercing an axionic domain wall embedded in AdS$_4$ along which $\theta$ changes. 
In this case, in contrast to the case of a constant $\theta$ parameter, the induced Aharonov-Bohm phases of the upper and lower boundary endpoint vortices/antivortices of a cosmic string do not cancel in a process in which one cosmic string is moved around another identical one. This is because in this case the Chern-Simons term on the upper hemisphere of the AdS boundary is induced with a different prefactor than the Chern-Simons term on the lower hemisphere and thus $|j^0_{(vortex)}| \neq |j^0_{(antivortex)}|$. The induced boundary Lagrangian can be written at the boundary $i$ ($i = 1,2$) as (\ref{bo}) 
\begin{equation}
\mathcal{L}_i = \frac{2}{e}\,\theta_i\, dA \wedge d\theta_{(v)} + \theta_i A\wedge dA
\label{lagrag}
\end{equation}
where $\theta_i$ are the values of the $\theta$ parameter at the $i$th boundary. In our convention, boundary number $1$ is the upper hemisphere of the conformal boundary of global AdS$_4$ and boundary number $2$ is the lower hemisphere. Both hemispheres are separated by the domain wall. This wall will show up in the boundary theory as a kink. However, as long as
the string pierces the domain wall and its endpoint vortices are well separated from the kink, we can integrate the above by parts and redefine $A$ in each hemisphere to yield
\begin{equation}
\mathcal{L}_i = -A \wedge \star j_{(vortex)} + \frac{e^2}{16\pi^2 \theta_i}  A\wedge dA
\label{lagrag2}
\end{equation} 

When we identify $\theta_i$ in (\ref{lagrag2}) with $\mu$ in (\ref{duallag})\footnote{Note that in (\ref{lagrag2}) the Lagrangian is expressed in terms of $A$ whereas in (\ref{duallag}) the dual language and the corresponding field $B$ is used. As mentioned already in the last paragraph of subsection \ref{subsec:3dknown}, instead of (\ref{duallag}) we could have obtained the same Lagrangian as (\ref{duallag}) with $B$ replaced by $A$ if we integrated out $B$ instead of $A$ in the discussion in subsection \ref{subsec:3dknown}. When we compare (\ref{lagrag2}) with (\ref{duallag}) we are obviously comparing it with (\ref{duallag}) where $B$ is replaced by $A$.}, this Lagrangian (\ref{lagrag2}) is nothing but minus the dual Lagrangian (\ref{duallag}) which describes the Chern-Simons vortices.\footnote{Even though the boundary part of the bulk Lagrangian (\ref{bo}) reproduces (\ref{duallag}), we must of course remember that the fields of (\ref{bo}) are inherently of (3+1) dimensions. The correspondence works because here we have a `radial' decomposition. In that setting, we are interpreting the bulk fields here as a trivial extension of their boundary counterpart (along the radial direction).} In the case of the constant $\theta$ parameter discussed above, $\theta_1 = \theta_2$ which induces boundary Lagrangians of the type (\ref{duallag}) with both having the same constant in front of the Chern-Simons term. This leads to the above mentioned cancelation of the induced Aharonov-Bohm phases of the upper and lower boundary vortices/antivortices of the cosmic string since in this case $j^0_{(vortex)} = -j^0_{(antivortex)}$. In the case of a non-constant $\theta$ parameter (e.g. which is shifted along the domain wall with $\theta_1 \neq \theta_2$), the induced Chern-Simons terms on the boundaries arise with different prefactors. Therefore, the charges of the upper endpoint vortices are different than the charges of the lower endpoint antivortices (for the same strings) and the Aharonov-Bohm phases do not cancel. In such a case, the upper and lower boundary endpoint (Chern-Simons) vortices/antivortices of the cosmic string taken together obey fractional statistics (as discussed in subsection \ref{subsec:4dtheta}) and the statistics in the bulk and on the boundary matches. In other words, we can say that the statistics of the bulk cosmic string can be obtained by considering only the statistics of the boundary endpoint vortices/antivortices of the string.

\subsection{Superfluid cosmic string and boundary vortices with additional current}\label{subsec:4d3daddgauge}

Let us now argue that the boundary vortices/antivortices of the (superconducting) cosmic string which we discussed in subsections \ref{subsec:4daddgauge} and \ref{subsec:wittstring} are vortices/antivortices of the kind we have discussed in subsection \ref{subsec:3daddcurrent}. We will again study the Aharonov-Bohm phase of the upper and lower boundary vortices/antivortices from both bulk and boundary perspectives in a process where one (superconducting) string is adiabatically taken around another identical one in such a way that the initial positions of the strings get exchanged.
\newline

The low energy dual Lagrangian with constant parameter $\theta$,
\begin{equation}
\mathcal{L} = -\frac{1}{2\pi} dB \wedge d\theta_{(v)} - \theta E \wedge dB\, ,
\end{equation}
can be written as
\begin{equation}
\mathcal{L} = \mathcal{L}_{bulk} + \mathcal{L}_{bdy} \, ,
\end{equation}
with
\begin{equation}
\mathcal{L}_{bulk} \equiv B\wedge\star j^{(vortex)} + \theta dE \wedge B \, ,
\label{llll}
\end{equation}
\begin{equation}
\mathcal{L}_{bdy} \equiv d \left( {\cal B}\wedge\star j_{(vortex)} - \theta E\wedge d{\cal B}\right) \, ,
\label{rrrr}
\end{equation}
where in the latter equation we have used that here $B = d{\cal B}$ on shell.
If we couple the current (\ref{currr}),
and its boundary current \eqref{eq:jbdy},
to (\ref{llll}) and (\ref{rrrr}) respectively, then the resulting bulk Lagrangian is the Lagrangian which we have considered in subsection \ref{subsec:4daddgauge} and the resulting boundary Lagrangian is equal to minus the Lagrangian which we have considered in subsection \ref{subsec:3daddcurrent} (when we identify $\theta$ in (\ref{rrrr}) with the parameter $- \kappa$ used in subsection \ref{subsec:4daddgauge}). Thus, in this sense, the vortices which we have considered in subsection \ref{subsec:3daddcurrent} can be viewed as the boundary vortices of the cosmic string which we have considered in subsection \ref{subsec:4daddgauge}.
\newline

If $\gamma(\sigma^0, \sigma^1 = \sigma^1|_{boundary\,1}) \neq \gamma(\sigma^0, \sigma^1 = \sigma^1|_{boundary\,2})$, the electric charges of the upper and lower boundary endpoint vortices/antivortices are different and induce different Aharonov-Bohm phase shifts which do not cancel. Also, as follows from the discussion in subsection \ref{subsec:3daddcurrent}, the phase shift induced on the boundary is the same as the one in the bulk (\ref{shi}). Let us here reemphasize that the boundary current \eqref{eq:jbdy} is precisely such that the total charge $Q \equiv \int_\Sigma \tilde{J}_{0,bulk} + \int_{\partial\Sigma} \tilde{J}_{0,bdy}$ vanishes, as discussed towards the end of section \ref{subsec:4daddgauge}. This is the origin of the matching statistical phase.

In summary, the conclusion that the Aharonov-Bohm phase shifts are equal to the analogous phase shifts of the cosmic strings in the bulk, applies equally for both setups which we have discussed.

\section{Summary and outlook}\label{sec:concl}

In this work, we have separately demonstrated (in two different setups) that electrically charged vortices in $(2 + 1)$ spacetime dimensions and electrically charged cosmic strings in $(3 + 1)$ dimensions obey fractional statistics. In both setups, we have explicitly calculated the induced Aharonov-Bohm phase shifts in processes in which two identical vortices or strings are rotated around each other. As we have mentioned throughout the text, some of these results are well known: e.g. as we discussed in subsection \ref{subsec:3dknown}, it is well known that electrically charged Chern-Simons vortices in $(2 + 1)$ dimensions obey fractional statistics \cite{Frohlich:1988qh} and it is also well known that (as we discussed in subsection \ref{subsec:4daddgauge}), cosmic strings in $(3 + 1)$ dimensions can obey fractional statistics if a certain additional current is localized on the string \cite{Aneziris:1990gm}. To our knowledge, the presentations which we gave in subsection \ref{subsec:3daddcurrent} and in particular in subsection \ref{subsec:4dtheta}, have however not appeared in the literature so far, although there are related works such as \cite{Hartnoll:2006zb}. In section \ref{sec:4d3drel} we combined the discussions of the previous two sections and presented a unified way of understanding the statistics of the cosmic strings in a $(3 + 1)$ dimensional spacetime with boundary and the statistics of corresponding boundary endpoint vortices/antivortices of the string which are located on the boundary of the spacetime. In both setups that we have considered, the cosmic strings obey fractional statistics if and only if their boundary endpoint vortices and antivortices carry different electric charges. In particular, in our parametrization, both the bulk and the boundary part of the currents are defined in such a way that even though they are not separately conserved, the combined total current is conserved as is expected for any consistent gauge theory. Our final result might be a very general criterion, not only applicable to the two ways of charging cosmic strings which we have considered explicitly. 
In other words: \textit{cosmic strings in spacetimes with boundary obey fractional statistics if and only if their boundary endpoint vortices and antivortices carry different electric charges.} Since the statistical phase shifts are purely due to the topological terms both at the bulk and on the boundary, it is clear that our result goes through for any suitable manifold $\mathcal{M}$ which can support these topological solutions.
\newline

This result might have generalizations to higher dimensional extended objects in higher spacetime dimensions with boundary. In fact, one can wonder under what conditions membranes in concrete theories can obey fractional statistics. Given our results, one can expect that e.g. two-dimensional membranes in five dimensional spacetimes with boundary obey fractional statistics if and only if their boundary endpoint strings carry different electric charges. 
\newline

Throughout our work, we have worked in the probe limit in which the backreaction of the topological objects on the spacetime is absent. Although, for a given spacetime it is not easy to determine the backreaction effects completely analytically, as this would require to solve the full coupled Einstein-Higgs equations, in certain approximations backreaction effects have been studied (for example in \cite{Dehghani:2001ft} for the case of cosmic strings in AdS). It might be interesting to study such backreaction effects in the context of fractional statistics which we have considered.
\newline

Our results can have several interesting applications in different contexts. We want to conclude our work by commenting on some of them.
First, as we have already mentioned several times in this work, our configurations can be naturally extended to global AdS spacetime, since AdS cosmic strings exist as solutions of (\ref{thetalag}) \cite{Dehghani:2001ft}. For us, it means that the fractionally charged cosmic strings are embedded in AdS spacetime with anyonic boundary endpoint vortices/antivortices located on the conformal boundary of AdS. In section \ref{sec:4d3drel}, we have already focused on such setups.
\newline

In the literature, e.g. in \cite{Dehghani:2001ft, Dias:2013bwa}, setups with (Nielsen-Olesen type) vortices located on the AdS boundary (which are endpoints of cosmic strings in the AdS bulk) have already been studied in the context of the AdS/CFT correspondence. In \cite{Dias:2013bwa} it has been emphasized that in the context of AdS/CFT (which relates a gravitational bulk theory to a conformal field theory on the AdS boundary), these lower dimensional vortices/antivortices can be understood as conformal defects (of the low energy field theory on the boundary). These defects break the full conformal group $SO(3,2)$ of the boundary field theory down to $SO(2,1) \times SO(2)$. So in this case, the boundary field theory is only invariant under the subgroup $SO(2,1) \times SO(2)$.
To our knowledge, the possible impact of fractionally charged anyonic vortices on such conformal defects has not yet been studied in the literature.
In this setting, it will thus be interesting to investigate this question both from the perspectives of a boundary vortex and also for the bulk string-vortex.
\newline

Our results can also have interesting applications at finite temperature and, in the context of AdS/CFT, will closely relate to the studies of holographic superconductors \cite{Hartnoll:2008vx} and to the studies of the fractional quantum hall effect \cite{KeskiVakkuri:2008eb, Nastase:2016niy}. Because vortices located at the AdS boundary have already been studied in such contexts in \cite{Montull:2009fe, Dias:2013bwa}, one might hope to learn the effects of fractional statistics on such condensed matter applications.
\newline

Finally, our results may have implications in the physics of Aharonov-Bohm type black hole hair. In fact, it is well known that black holes can be charged under discrete $Z_N$ symmetry \cite{Coleman:1991ku}, and in those cases, cosmic strings do appear as solutions. It is therefore an interesting question as to whether our studies on the fractional statistics of cosmic strings might have some implications on the physics of hairy black holes, in particular in the context of holography \cite{Montull:2011im}.

\vspace{1.5 cm}
\centerline{\bf Acknowledgements}
\vspace{0.5 cm}
\noindent
We thank Nabil Iqbal for valuable discussions and Uwe-Jens Wiese for his detailed comments on the draft. The work of DS is funded by the NCCR SwissMAP (The Mathematics of Physics) of the Swiss Science Foundation. NW acknowledges support by FNU grant number DFF-6108-00340.

\end{document}